\documentclass[12pt,preprint]{aastex}

\def\ltsima{$\; \buildrel < \over \sim \;$}
\def\gtsima{$\; \buildrel > \over \sim \;$}
\def\lsim{\lower.5ex\hbox{\ltsima}}
\def\gsim{\lower.5ex\hbox{\gtsima}}
\def\lapp{\ifmmode\stackrel{<}{_{\sim}}\else$\stackrel{<}{_{\sim}}$\fi}
\def\gapp{\ifmmode\stackrel{>}{_{\sim}}\else$\stackrel{<}{_{\sim}}$\fi}

\newcommand{\masyr}{${\rm mas\, yr^{-1}}$}

\usepackage{amsmath}
\usepackage{textcomp}
\usepackage{amssymb}
\usepackage{multirow}
\usepackage{booktabs}
\usepackage{graphicx}

\newdimen\minuswidth    
\setbox0=\hbox{$-$}
\minuswidth=\wd0
\catcode`@=\active
\def@{\kern\minuswidth}
\setbox0=\hbox{\rm0}
 
\shorttitle{} 
\shortauthors{Massari et al.}
 
\begin{document} 
\title{Proper motions in Terzan~5: membership of the multi-iron sub-populations and first constrain to the orbit}

\author{
D. Massari\altaffilmark{1,2},
E. Dalessandro\altaffilmark{3},
F.R. Ferraro\altaffilmark{3},
P. Miocchi\altaffilmark{3},
A. Bellini\altaffilmark{4},
L. Origlia\altaffilmark{1},
B. Lanzoni\altaffilmark{3},
R. M. Rich\altaffilmark{5},
A. Mucciarelli\altaffilmark{3}
}
\affil{\altaffilmark{1}INAF-Osservatorio Astronomico di Bologna, via
  Ranzani 1, 40127, Bologna, Italy} 
\affil{\altaffilmark{2}Kapteyn Astronomical Institute, University of Groningen, 
PO Box 800, 9700 AV Groningen, The Netherlands}   
\affil{\altaffilmark{3}Dipartimento di Fisica e Astronomia, Universit\`a degli Studi
di Bologna, v.le Berti Pichat 6/2, I$-$40127 Bologna, Italy}
\affil{\altaffilmark{4}Space Telescope Science Institute, 3700 San Martin
  Drive, Baltimore, MD 21218, USA}
\affil{\altaffilmark{5}Department of Physics and Astronomy, Math-Sciences 8979, 
UCLA, Los Angeles, CA 90095-1562, USA} 
  
\altaffiltext{$^\ast$}{Based on observations (GO12933, GO9799) with the
  NASA/ESA \textit{Hubble Space Telescope}, obtained at the Space
  Telescope Science Institute, which is operated by AURA, Inc., under
  NASA contract NAS 5-26555.}
\date{9 July, 2015}

\begin{abstract}

By exploiting two sets of high-resolution images obtained with HST
ACS/WFC over a baseline of $\sim 10$ years we have measured relative
proper motions of $\sim 70,000$ stars in the stellar system
Terzan~5. The results confirm the membership of the three sub-populations 
with different iron abudances discovered in the system.  The orbit of the
system has been derived from a first estimate of its absolute proper
motion, obtained by using bulge stars as reference. 
The results of the integration of this orbit within an axisymmetric Galactic
model exclude any external accretion origin for this cluster.
Terzan~5 is known to have chemistry similar to the Galactic bulge;
our findings support a kinematic link between the cluster and the bulge, 
further strengthening the possibility that Terzan~5 is the fossil remnant 
of one of the pristine clumps that originated the bulge.

\end{abstract}
 

\section{INTRODUCTION}
Terzan 5 is a stellar system located at the edge of the inner bulge of the Galaxy
($l=3.8395$\textdegree, $b=1.6868$\textdegree) at a distance of 5.9 kpc from the Sun
(\citealt{valenti07}). Because of the large  and spatially variable extinction in that
region of the sky (\citealt{massari}), observations of this system are extremely
challenging, particularly in the optical bands.  This is why the complex nature of Terzan
5 has been revealed only recently by means of high resolution IR photometric
(\citealt{f09}) and spectroscopic (see \citealt{origlia})  observations of its stellar
populations with the ESO-VLT and Keck telescopes. These studies revealed that Terzan 5
hosts two distinct populations with significantly different iron content
($\Delta$[Fe/H]$=0.5$ dex), different level of $\alpha$-element enhancement, and no
evidence of the anti-correlation among light elements commonly observed in globular
clusters (GCs; e.g., \citealp{carretta14} and references therein). The radial
distribution of the two populations is incompatible with that of field stars. In
addition, the metal-rich component has been found to be significantly more centrally
concentrated than the metal poor one \citep{l10}. Also, the two populations have been found
to share the same mean radial velocities (\citealp{massari14a} and references therein) 
and the same center of gravity \citep{l10}. All these features add weight to the membership of these
populations. However, even with the spatial distribution and radial velocity membership
secure, their actual membership has been questioned (e.g. \citealt{willman}) because of
the strong field star contamination affecting many GCs in the direction of the bulge (see
\citealt{valenti07, valenti10}).   Recently, also a third, metal-poor and
$\alpha$-enhanced population has been discovered (\citealt{o13}).  Although small
in number, such a component shares the same systemic radial velocity of the
cluster (\citealt{o13}) and it has survived to all statistical decontamination tests (see the
discussion in \citealt{massari14b}).

In order to properly separate genuine Terzan 5 stars from foreground and background
sources, we measured high-precision relative proper motions (PMs) of individual stars in
the direction of the system, by exploiting the superb astrometric accuracy of the {\it
Hubble Space Telescope} (see e.g. \citealt{jay10,bellini14, 
watkins15,massari13, dalex13}, and references therein). Here we report on the results
of this proper motion study, which also confirms the membership of the three iron groups.

This paper is organized as follows. In
Section \ref{datamoti} we present the used data set, while in Section \ref{motirel} we
describe the techniques adopted to measure PMs. Finally, we present the results of
our work in Section \ref{results} and we summarize the conclusions in Section
\ref{conclusions}.

\section{OBSERVATIONS AND DATA REDUCTION}
\label{datamoti}
In the context of a multi-wavelength program aimed at studying the
photometric properties of the multi-iron sub-populations of Terzan 5,
we have acquired (GO-12933, PI: Ferraro) a set of optical and
near-infrared (IR) images of the system, by using the Wide Field
Channel (WFC) of the Advanced Camera for Surveys (ACS) and the IR
channel of the the Wide Field Camera 3 (WFC3) onboard the Hubble Space
Telescope (HST).  In the following we will focus on the optical data set
only, since the subject of the present paper is the accurate measure
of the stellar PMs, and  the larger pixel scale 
of the WFC3 IR camera detector
($\sim0.13 \arcsec\,$pixel$^{-1}$) makes the secured IR images 
unsuitable for this purpose.  The
ACS/WFC is made up of two $2048 \times 4096$ pixel detectors with a
pixel scale of $\sim0.05 \arcsec\,$pixel$^{-1}$ and separated by a gap
of about 50 pixels. The total field of view (FoV) is $\sim200\arcsec
\times 200\arcsec$.  The optical data set consists of $5\times365$ s
images in F606W, $5\times365$ s images in F814W, and one short
exposure per filter ($50$ s and $10$ s, respectively). The
observations were performed on August 18th, 2013 and provided an
optimal second-epoch data set for PM measures.
  
The first epoch images (GO-9799, PI: Rich) consist of two deep (340 s)
exposures in the F606W and F814W filters, and one short exposure (10
s) in the F814W filter, also acquired with the ACS/WFC. This data set
was used to construct the deepest optical color-magnitude diagram
(CMD) of Terzan5 (see \citealt{f09, l10}). The images were
acquired on September 9th, 2003.  Thus, the combination of the two
optical data sets provides a total time baseline of $\sim 9.927$ yrs.

\section{RELATIVE PROPER MOTIONS}
\label{motirel}
 The analysis has been performed on \texttt{$\_$flc} images, which have been
flat-fielded, bias-subtracted and corrected for Charge Transfer
Efficiency (CTE) losses by the standard HST calibration pipeline
(CALACS) adopting the
pixel-based correction described in \cite{jaybedin10} and
\cite{ubedajay}.  The procedure used to derive relative PMs is
described in detail in \cite{massari13}. Here we provide only a brief
description of the main steps of the analysis.  The first step
consists in the photometric reduction of each individual exposure of
the two epochs with the publicly available program
\texttt{img2xym$\_$WFC.09$\times$10} \citet{jayking06, jay08}.  This program
uses a  filter-dependent library of spatially varying PSF models plus a single
time-dependent perturbation PSF to account for focus changes or
spacecraft breathing. The final output is a catalog with instrumental
positions and magnitudes for a sample of sources above a given flux
threshold in each exposure.  Star positions were then corrected in
each catalog for geometric distortion, by means of the solution
provided by \cite{jayacs}.

\begin{figure}[!htb]
 \centering%
 \includegraphics[scale=0.6]{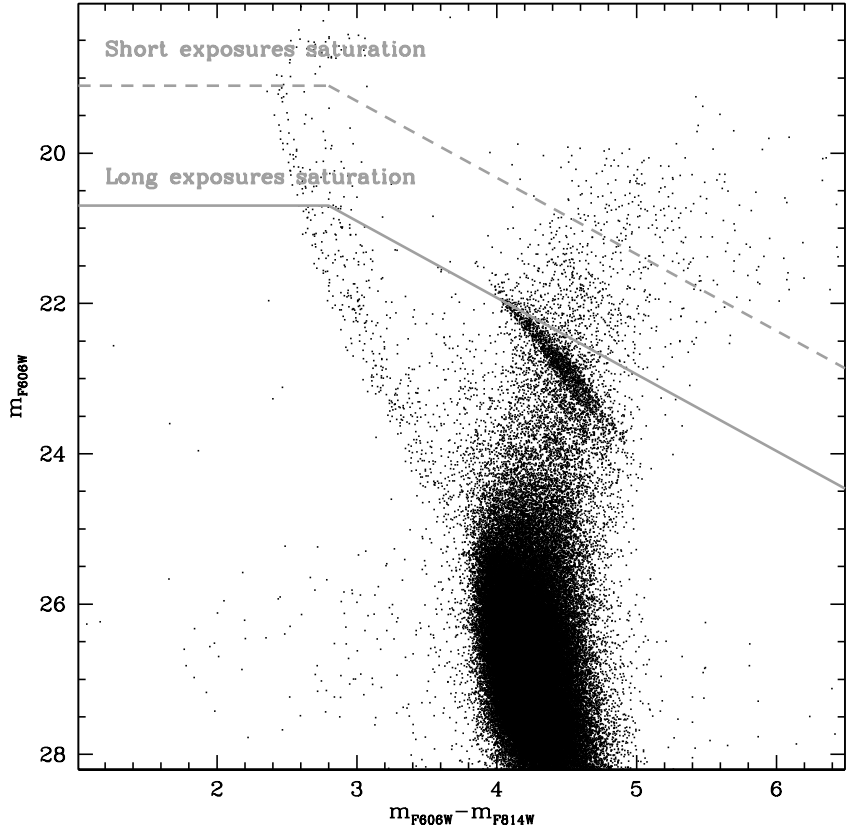}
 \caption{\small ({\it m}$_{{\rm F606W}}$, {\it m}$_{{\rm
       F606W}}-${\it m}$_{{\rm F814W}}$) CMD of Terzan 5. All the
   cluster evolutionary sequences are broadened because of
   differential reddening effect. A bright, blue sequence is clearly
   separated from the cluster sequences and it is likely composed of
   young field stars. The magnitude saturation limits of the short and
   long exposures are marked with the dashed and solid lines,
   respectively.}
\label{vvimoti}
\end{figure}

Figure \ref{vvimoti} shows the ($m_{\rm
  F606W}$, $m_{\rm F606W}- m_{\rm F814W}$) CMD of Terzan 5 using the
second epoch data set (Fig. \ref{vvimoti}). In particular, for stars
fainter than $m_{\rm F606W}\simeq20.7$ mag, which is the saturation
limit of the deep images (solid grey line in Fig. \ref{vvimoti}), all
stars detected in at least 3 (out of 5) deep single-exposures per filter are
plotted.  Following the standard procedure (\citealt{lanzoni07,dale08}),
for each star, the magnitudes obtained from each single
exposure in each filter were first homogeneized, then averaged. The
mean and standard deviation were finally adopted as instrumental
magnitude and photometric error, respectively (see
\citealt{ferraro91,ferraro92}).  For $m_{\rm F606W}<20.7$ mag, we
considered only stars measured in both the short exposures, which
saturate at $m_{\rm F606W}\simeq19.1$ mag (dashed grey line in
Fig. \ref{vvimoti}).  The instrumental magnitudes have been calibrated
onto the Johnson photometric system using the stars in common with the
catalog of \cite{l10}. As apparent from the figure, the evolutionary
sequences of Terzan 5 are clearly distinguishable although strongly
affected by differential reddening.  The main sequence extends for
almost $4$ magnitudes below the turn off.  A blue sequence is visible
at $m_{\rm F606W}<24.5$ mag and ($m_{\rm F606W}- m_{\rm F814W})<3.6$
  mag and it remains well separated from the cluster RGB. This
  sequence is likely populated by young disk stars.

The second step in determining relative PMs is to astrometrically
relate each exposure to a distortion-free reference frame, which from
now on we will refer to as the {\it master frame}.  Since no
high-resolution astrometry other than that coming from these data sets
is available, we defined as master frame the catalog obtained from the
combination of all the second-epoch single-exposure catalogs corrected
for geometric distortion. In this way, the master frame is composed
only of stars with at least $10$ position measurements (5 for each
filter).  We then applied a counter-clockwise rotation of
$91.163$\textdegree~ in order to give to the master frame the same
orientation as the absolute reference frame defined by the Two Micron
All Sky Survey (2MASS) catalog (see \citealt{l10}).  We then
transformed the measured position of each star in each exposure into
the {\it master frame} by means of a six-parameter linear
transformation.  
In order to maximizes the accuracy, the transformations 
have been computed using only high signal-to-noise
({\it m}$_{{\rm F606W}}<24$) and unsaturated sources that could 
be considered likely cluster
members according to their position in the CMD.  
Also, in order to minimize the effects of CTE- and distrorsion-correction residuals
we treated each chip  
separately.  At the end of the procedure, for each star we have up to
 3 first-epoch position measurements and up to 12
second-epoch positions on the master frame.   
To estimate the relative PM of each star we
computed the median positions in the first and in
the second epoch by applying a 3$\sigma$-clipping algorithm. The
difference between the two median positions gives the star's  
displacement over $\Delta {\rm T}=9.927$ years.  Since 
for stars brighter
than {\it m}$_{{\rm F606W}}\simeq20.7$  one first-epoch and two
second epoch positions were available, we adopted the
single position measured in the first epoch and the mean between the
two positions measured in the second epoch.  The errors in each
direction and within each epoch were computed as:
\begin{equation}
\sigma_{1,2}^{{\rm X,Y}} = \frac{{\rm rms_{1,2}^{{\rm X,Y}}}}{\sqrt{N_{1,2}}},
\end{equation}
where rms$_{1,2}$ is the rms of the positional residuals about the
median value, and N$_{1,2}$ is the number of measurements. Therefore,
the error in each PM-component associated to each star is simply the
sum in quadrature between first- and second-epoch
errors:\ $\sigma_{\rm PM}^{{\rm X}}=\sqrt{(\sigma_1^{{\rm
      X}})^{2}+(\sigma_2^{{\rm X}})^{2}}/\Delta {\rm T}$ and
$\sigma_{\rm PM}^{{\rm Y}}=\sqrt{(\sigma_1^{{\rm
      Y}})^{2}+(\sigma_2^{{\rm Y}})^{2}}/\Delta {\rm T}$.  The error
associated to the PM of the brightest stars measured only in the short
exposures were computed by adopting as positional uncertainties the
typical errors determined in the long exposure catalogs at the same
{\it instrumental} magnitude.

\begin{figure}[!htb]
 \centering%
 \includegraphics[scale=0.45]{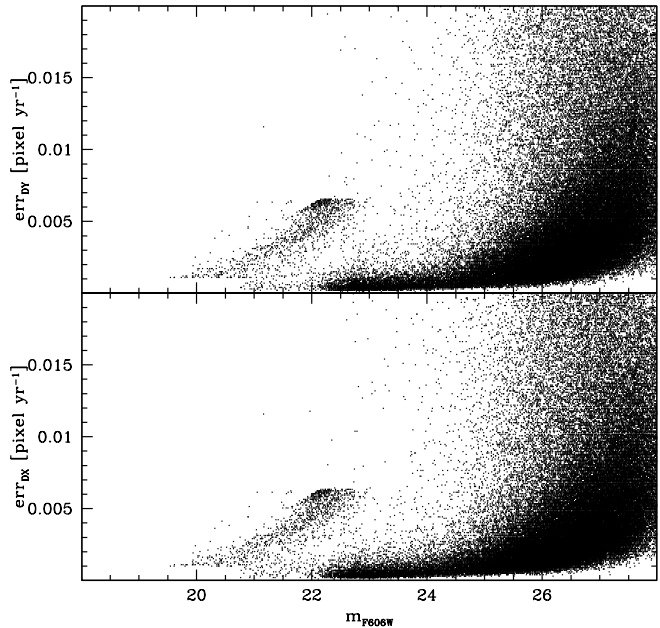}
 \caption{\small Uncertainties in the X and Y displacements in units
   of pixel yr$^{-1}$.  Bright, well measured stars have displacement
   errors typically smaller than 0.007 pixel yr$^{-1}$ in each
   coordinate. The separated sequence at larger errors which arises
   for {\it m}$_{{\rm F606W}}<22.5$ is made up of stars measured only
   in the short exposures. }
\label{errmoti}
\end{figure}

We repeated the entire procedure several times, in order to reach a
stable number of stars in the reference list. In performing these
iterations, stars were selected on the basis of the relative PMs
obtained in the previous step.  To be conservative, we decided to
build the final PM catalog taking into account only the $69,425$
unsaturated stars measured in at least 3 long exposures per epoch and
the $899$ stars brighter than the deep image saturation limit
measured in all the short exposures which have a total uncertainty
on the displacement smaller than 0.2 pixels. All the sources which saturate in
the latter images have been excluded.  The typical error as a function
of magnitude is shown in Figure \ref{errmoti}.  For well-exposed stars
it is smaller than 0.007 pixel yr$^{-1}$ in each coordinate.  Faint
stars or stars with only few measurements show larger errors, still
typically smaller than $0.02$ pixel yr$^{-1}$ (see the separated
sequence in Fig. \ref{errmoti}).

\subsection{The Vector Point Diagram}

\begin{figure}[!htb]
 \centering%
 \includegraphics[scale=0.5]{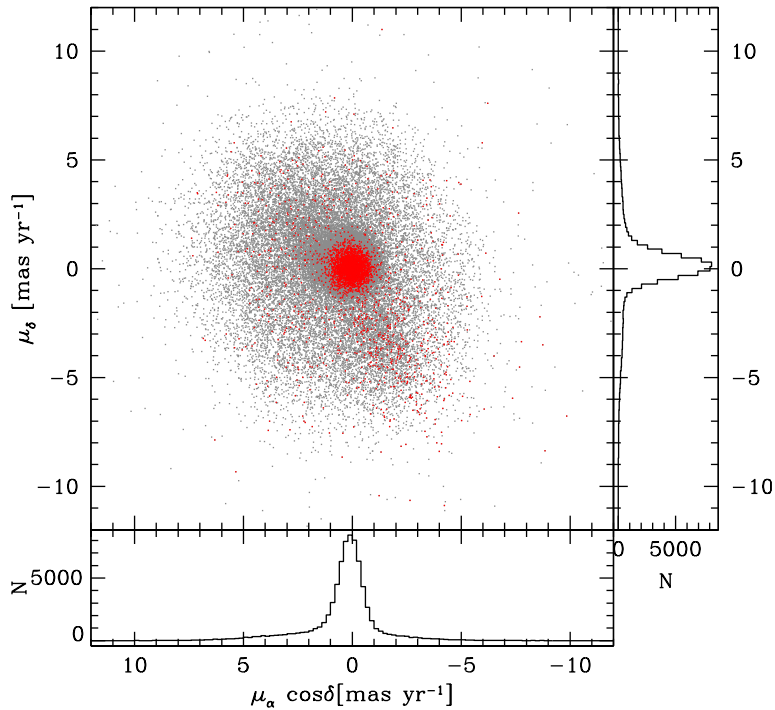}
 \caption{\small Vector point diagram (VPD) of the PMs measured for
   $70,324$ stars (black dots) in the direction of Terzan 5. Their
   distribution in the Right Ascension and Declination PM-component
   axes is shown in the histograms in the bottom and right panel,
   respectively. PMs measured for unsaturated stars with $m_{\rm
     F606W}<24$ mag are shown as red dots. At least two components are
   visible: the first showing a symmetric distribution centered around
   the origin and an asymmetric structure roughly centered at (-1.8,
   -3.5) \masyr. }
\label{vpdmoti}
\end{figure}

\begin{figure}[!htb]
 \centering%
 \includegraphics[scale=0.5]{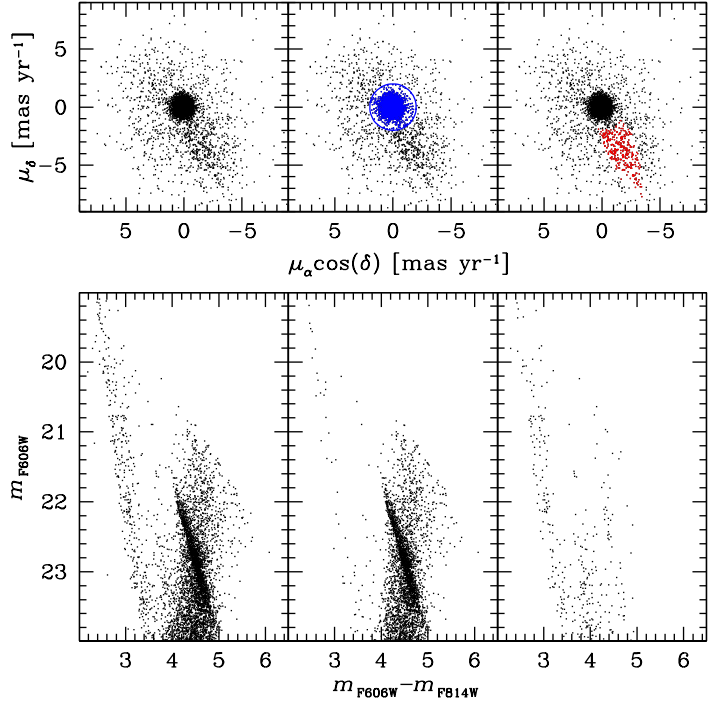}
 \caption{\small {\it Upper panels}: VPD of all stars brighter than
   m$_{F606W}=24$ mag.  The PMs of stars belonging to the symmetric
   component of likely members are highlighted in blue in the central
   panel, those of stars belonging to the asymmetric component
   centered at ($-1.8$, $-3.5$) \masyr\ are plotted in red in the
   right-hand panel. {\it Lower panels}: CMDs described by the
   PM-selected stars in the corresponding VPDs. When likely Terzan 5
   members are selected, only the cluster evolutionary sequences are
   visible, with the exception of few residual contaminating stars.
   On the other hand, the population belonging to the asymmetric
   component appears to be dominated by young foreground disk stars
   along the blue plume. }
\label{two_moti}
\end{figure}

\begin{figure}[!htb]
 \centering%
 \includegraphics[scale=0.45]{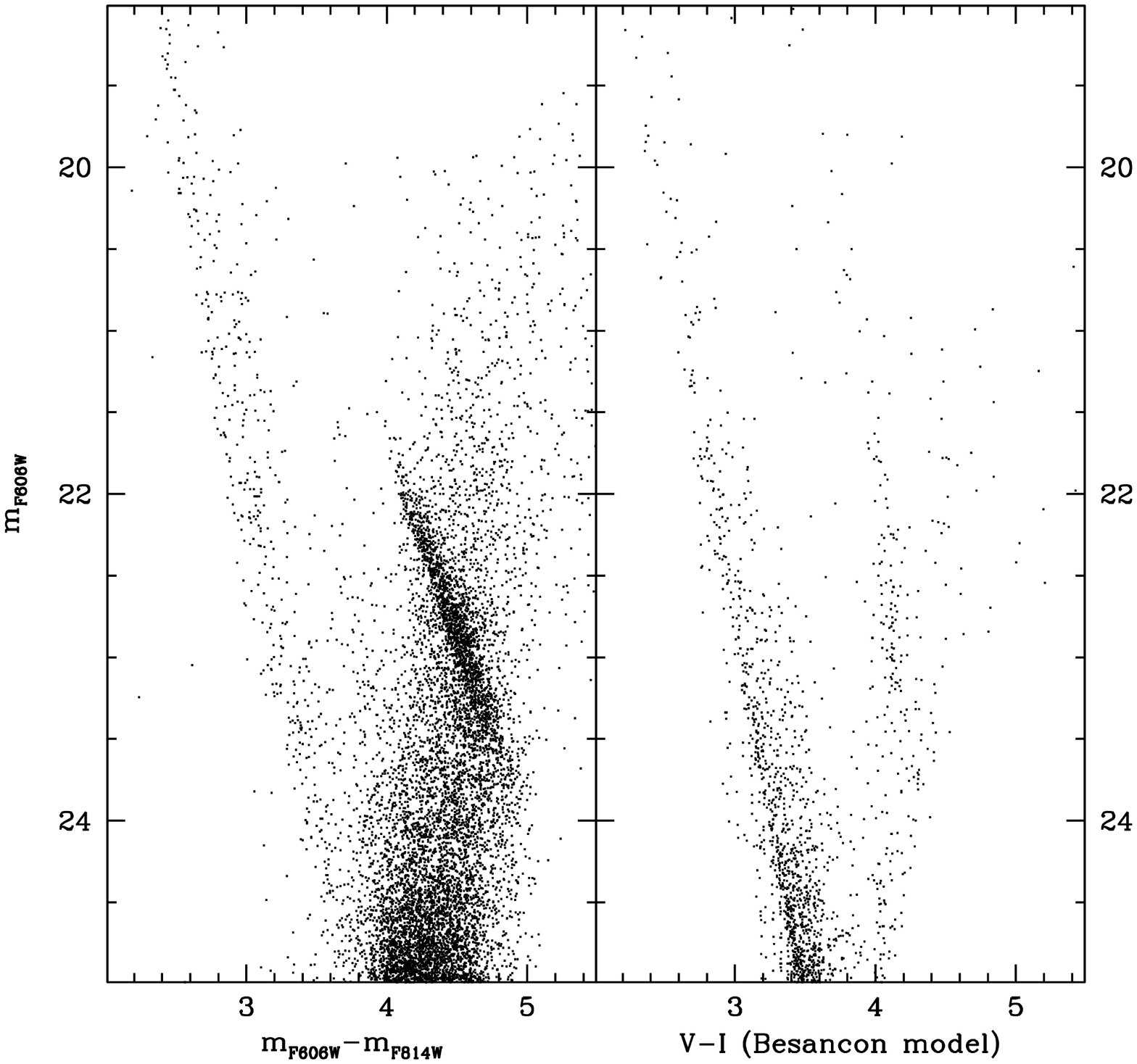}
 \caption{\small Comparison between the observed ACS CMD and that
   predicted by a simulation of the Besan\c{c}on model including only
   Galactic disk stars younger than 7 Gyr. Such a comparison clearly
   demonstrates that the bluer sequence in the CMD of Terzan 5,
   already identified as composed of field stars by the measured PMs,
   correspond to the MS of foreground disk stars. } 
\label{vvibes}
\end{figure}

We converted the PMs into units of \masyr by multiplying the measured
displacements by the pixel scale of the master frame ($0.05
\arcsec/$pixel) . Since the master frame has been oriented according
to the equatorial coordinate system, the X PM-component corresponds to
that projected along (negative) Right Ascension
($-\mu_{\alpha}\cos\delta$), while the Y PM-component corresponds to
that along Declination ($\mu_{\delta}$).  The output of this analysis
is summarized in Figure \ref{vpdmoti}, where we show the vector point
diagram (VPD) for all the stars of the final PM catalog.  The first
clear feature is that more than 70\% of the stars are distributed
within the innermost 1.5 \masyr\ from the origin of the VPD (see
histograms in Fig. \ref{vpdmoti}). The remaining fraction of stars
describes a sparser and asymmetric distribution out to about 10
\masyr.  In order to highlight these features, we selected only stars
with {\it m}$_{{\rm F606W}}<24$, which typically have the most
accurate PMs, and plotted them as
red points in Figure \ref{vpdmoti}.  Their distribution in the VPD
clearly shows at least two components. One is a symmetric distribution
centered around the origin, with PM weighted mean values
$\mu_{\alpha}\cos\delta=0.02\pm0.02$ \masyr\ and
$\mu_{\delta}=-0.01\pm0.02$ \masyr.  The other is an asymmetric
structure approximately centered around the coordinate
($\mu_{\alpha}\cos\delta\sim-1.8$, $\mu_{\delta}\sim-3.5$) \masyr\ in
the VPD.  The location of these two components in the CMD clearly
reveal their nature (see Figure \ref{two_moti}).  In fact, the stars
of the first component (shown as blue dots in the VPD) describe the
cluster evolutionary sequences in the CMD (central lower panel), with
a small degree of contamination left. On the other hand, the stars
belonging to the asymmetric component (red dots) correspond to the
blue plume in the CMD (right-lower panel), which is essentially
populated by young disk stars in the foreground of Terzan 5.  This is
further confirmed by the comparison with the prediction of the
Besan\c{c}on model (\citealt{robin}) with only young (t$_{age}<7$ Gyr)
Galactic disk stars for a field centered at the coordinates of Terzan
5 and covering the same FoV as that of the ACS/WFC (see Figure
\ref{vvibes}).

\subsection{Check for systematic errors}

Many systematic uncertainties may affect the measure of relative PMs
(see \citealt{bellini14} for a detailed description). It is therefore
important to check that the measured PMs of our final catalog do not
suffer from such systematic effects. Since we have only two epochs of
observations and a relatively small number of images per epoch (only
two long exposures were acquired in the first epoch) the full control
of all the possible systematics is not achievable. Nonetheless, we
have carefully checked for the presence of any systematics that our
data sets allow us to verify.

\begin{figure}[!htb]
 \centering%
 \includegraphics[scale=0.5]{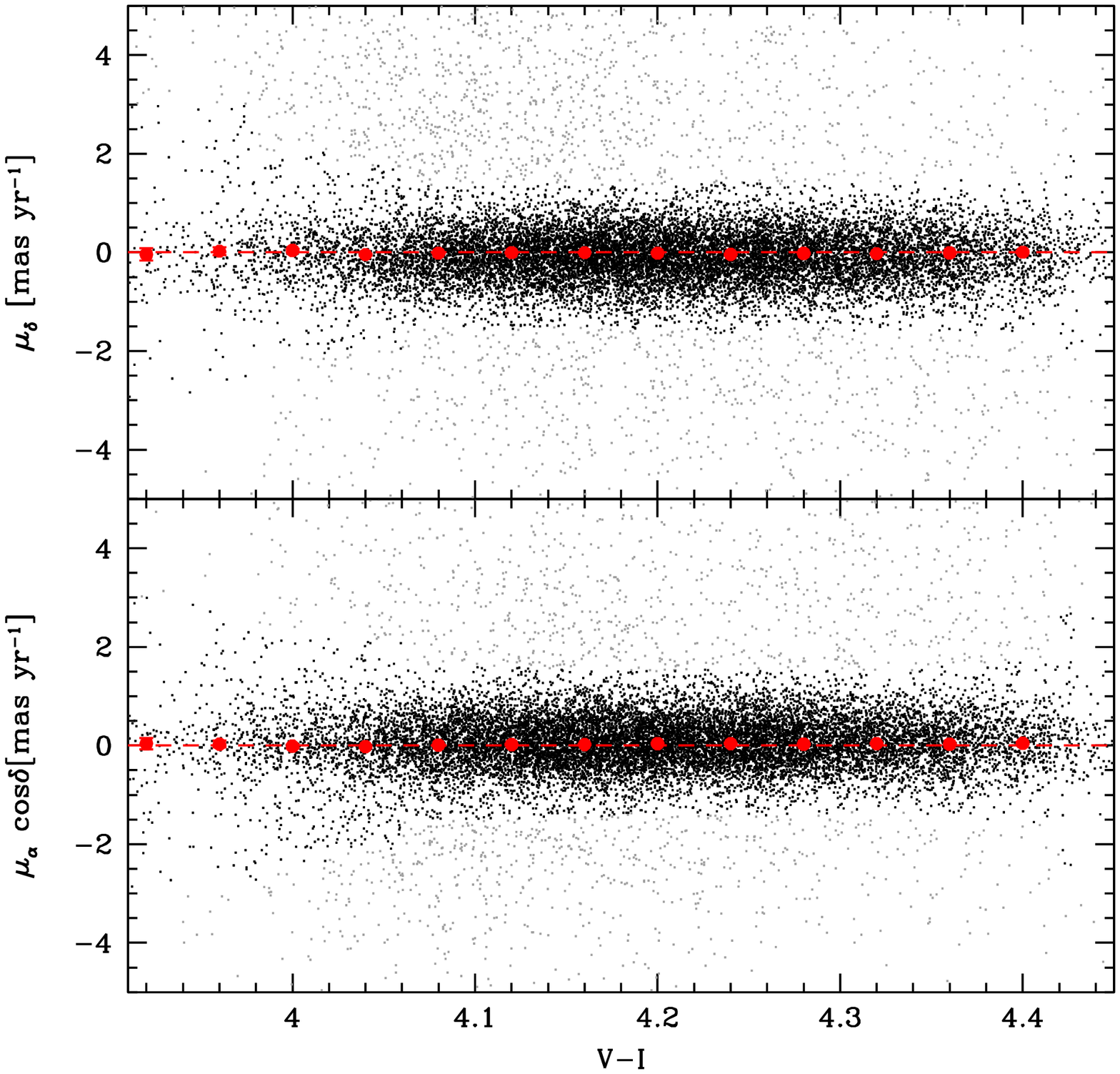}
 \caption{\small {\it Upper panel:} Y-component of the relative PMs vs
   observed colors. Red filled circles represent the 3-$\sigma$
   clipped mean PMs measured in color bins of $0.04$ magnitudes. Stars
   rejected by the clipping algorithm are shown in grey. All the mean
   values lie on the no-correlation line (red dashed line), thus
   ecluding any chromatic-induced systematics. {\it Lower panel:} same
   as above, for the PM X-component. }
\label{col_trend}
\end{figure}

First of all we looked for any possible chromatic-induced systematics
by checking that no trend between PMs and color exists for our sample.
Figure \ref{col_trend} shows the results of this test. By selecting
stars in a wide magnitude range ($23<${\it m}$_{{\rm F606W}}<26$), we
computed the 3-$\sigma$ clipped average value of our PMs (for both the
spatial components) in color bins of $0.04$ mag (red filled
circles). The errorbars are within the size of the symbols.  Clearly,
these values describe a flat relation (shown as a red dashed
line). The best least square linear fit to these data gives a null
angular coefficient within the uncertainty for both the PM components.
Thus, we can safely conclude that no trend with color exists for our
measurements.

\begin{figure}[!htb]
 \centering%
 \includegraphics[scale=0.5]{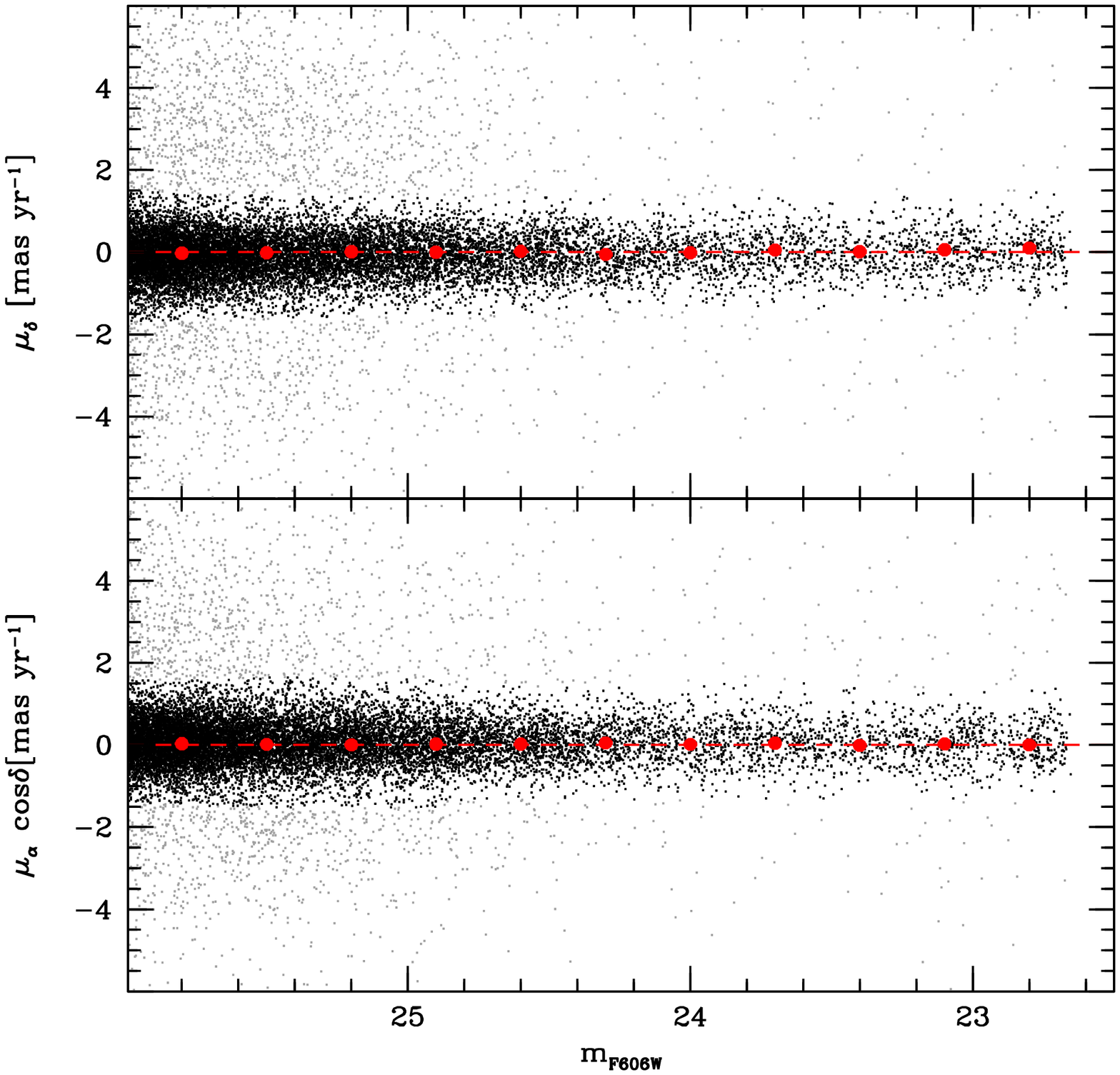}
 \caption{\small {\it Upper panel:} same plot as in Figure
   \ref{col_trend}, but in terms of observed magnitude.  Mean values
   have been computed in bins 0.3 mag wide. Also in this case, no
   trend is found between PMs and magnitude.  {\it Lower panel:} same
   as above, for the X-component of PMs. }
\label{mag_trend}
\end{figure}

We repeated a similar test looking for possible trends with the
observed magnitude in the range $19<m_{\rm F606W}<26$, thus to exclude
extremely faint sources that have intrinsically larger PM
errors. Figure \ref{mag_trend} shows the result of this test. Also in
this case, the best linear fit to the 3-$\sigma$ clipped average PM
values computed in bins of 0.3 mag is compatible (within the errors)
with a null angular coefficient line, thus demonstrating that no significant trend
is found also with magnitude.

Finally, we also carefully checked the existence of spurious
trends with the location of stars on the
detector. To do so, we followed the method described in \cite{bellini14},
by plotting maps of the measured PMs as a function of the position on the sky, 
where to each star we associated the average motion of the closest 100 neighbors. 
Overall we found a homogeneous distribution in both the PM components.

The tests discussed above demonstrate that the derived PMs do not
suffer from significant systematic effects. Also, the use of available
CTE corrected images for both epochs (see Sect.\ref{motirel}) should
have minimized the effect of CTE losses on the PM estimate. While we
cannot exclude that other systematics can affect out measurements, we
can firmly conclude that these are the best PMs we can measure with
the available data set.

\section{RESULTS}
\label{results}

\begin{figure}[!htb]
 \centering%
 \includegraphics[scale=0.5]{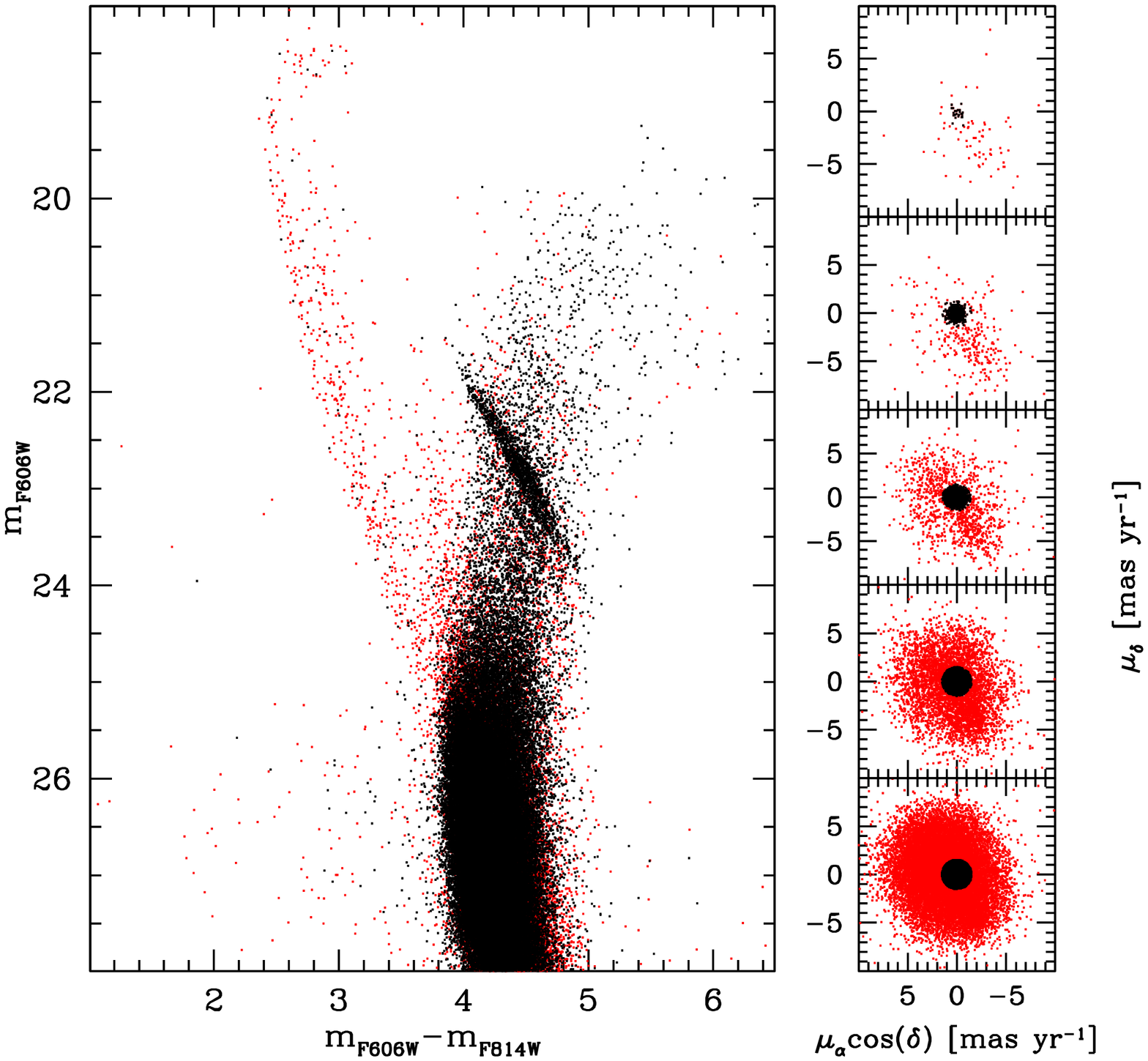}
 \caption{\small {\it Left panel:} optical CMD of Terzan 5, with
   likely member stars shown as black dots, and field sources
   (excluded by the PM-based selection) plotted in red.  The measured
   relative PMs are efficient in decontaminating the CMD and only
   cluster evolutionary sequences survive the selection
   criterion. {\it Right panels:} magnitude-binned VPDs for all the
   stars in the optical catalog. Each bin has a size of 2 mag. Sources
   are color coded as in the left panel. }
\label{vviclean}
\end{figure}

We used the measured relative PMs to ``clean'' the optical ($m_{\rm
  F606W}$, $m_{\rm F606W}-m_{\rm F814W}$) CMD of Terzan 5 described in
Section \ref{datamoti}. We defined as likely member stars all the
sources located within a distance smaller than 1.5 \masyr\ in the VPD
plane. As shown in Figure \ref{vpdmoti}, this appears to be a
reasonable assumption, since the bulk of the stars lie within this
limit. The CMD obtained from such a selection is shown in Figure
\ref{vviclean}.  Member stars are plotted as black dots, while non
members stars are shown in red. The selection applied leaves in the
CMD only stars clearly belonging to the cluster evolutionary
sequences, while excluding most of the outliers. A small degree of
contamination is still present because the distribution of field stars
(mainly bulge stars) in the VPD overlaps that of Terzan 5 members.
However, we can conclude that the PM analysis performed is efficient
in decontaminating the CMD from foreground and background sources.
We can therefore use this selection to assess the membership of the
sub-populations discussed by \cite{f09,origlia,o13,massari14b}.

\subsection{MAD Infrared CMD}

\begin{figure}[!htb]
 \centering%
 \includegraphics[scale=0.5]{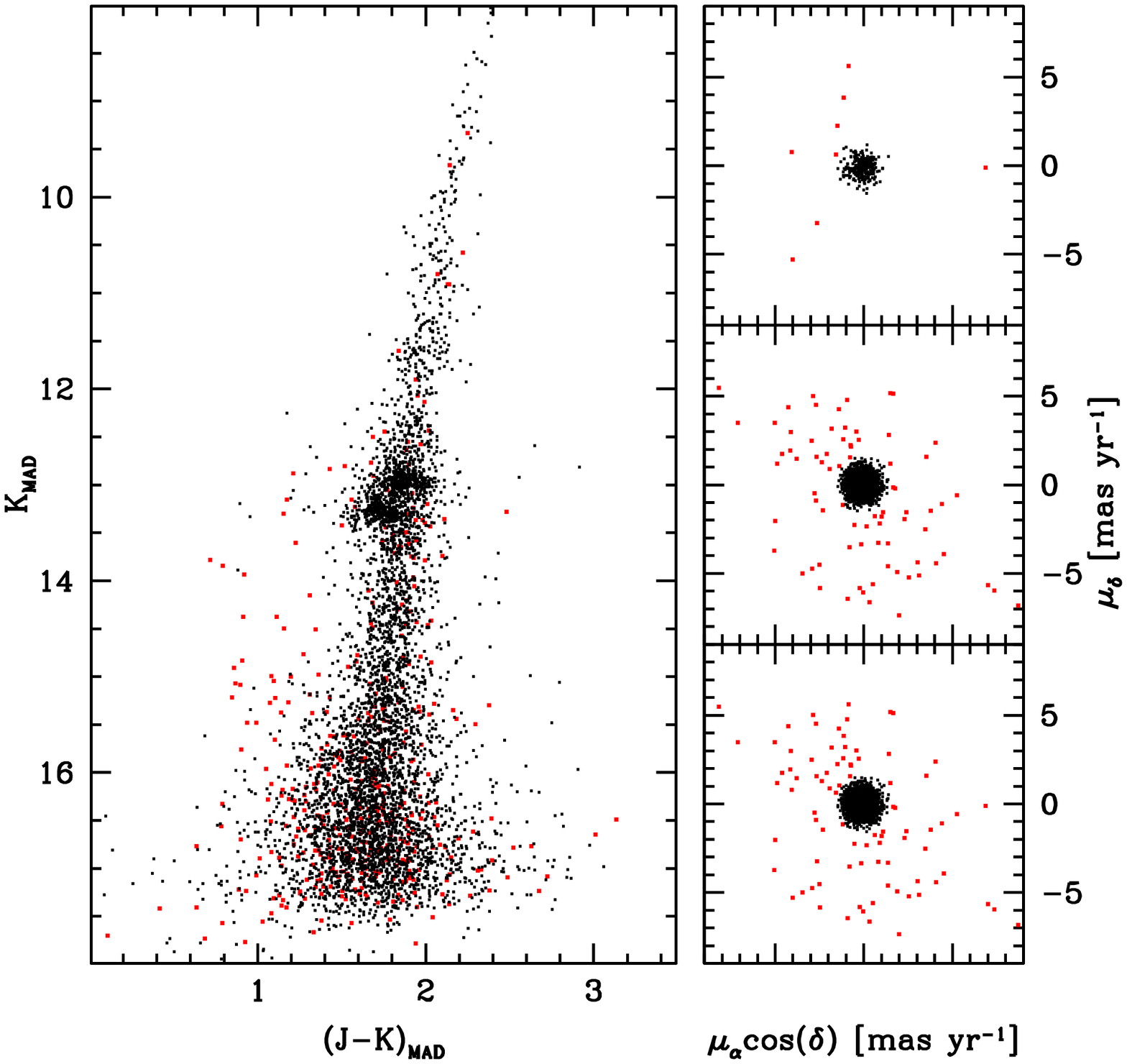}
 \caption{\small {\it Left panel:} IR CMD of Terzan 5 obtained from
   MAD observations. The PM-selected member stars are shown as black
   dots, while the sources having discordant PMs are marked in red.
   The presence of two distinct red clumps is well evident even after
   the PM-based selection. {\it Right panels:} magnitude-binned VPDs
   (each magnitude bin has a size of 3 mag). Stars are color-coded as
   in the left panel. }
\label{kjkclean}
\end{figure}

The two major sub-populations in Terzan 5 were identified in the ($K,
J-K$) CMD by \citet{f09}. Their radial distributions are incompatible
with that expected for background field stars, and the measured line
of sight velocities are consistent with the systemic radial velocity
of Terzan 5. However, the membership of the two populations has been
questioned (see, e.g., \citealt{willman}). To address this concern and to
assess their cluster membership from proper motions, we 
cross-correlated the MAD catalog with the PM catalog.  
The result is summarized in Figure \ref{kjkclean}.  The left
panel shows the IR CMD of Terzan 5 after the PM-based membership
selection (black dots). The sources recognized as contaminating field
stars based on the value of their PMs are also plotted in red.  The
decontaminated CMD clearly exhibits the two red clumps (RCs)
originally discovered by \citet{f09}.

\begin{figure}[!htb]
 \centering%
 \includegraphics[scale=0.5]{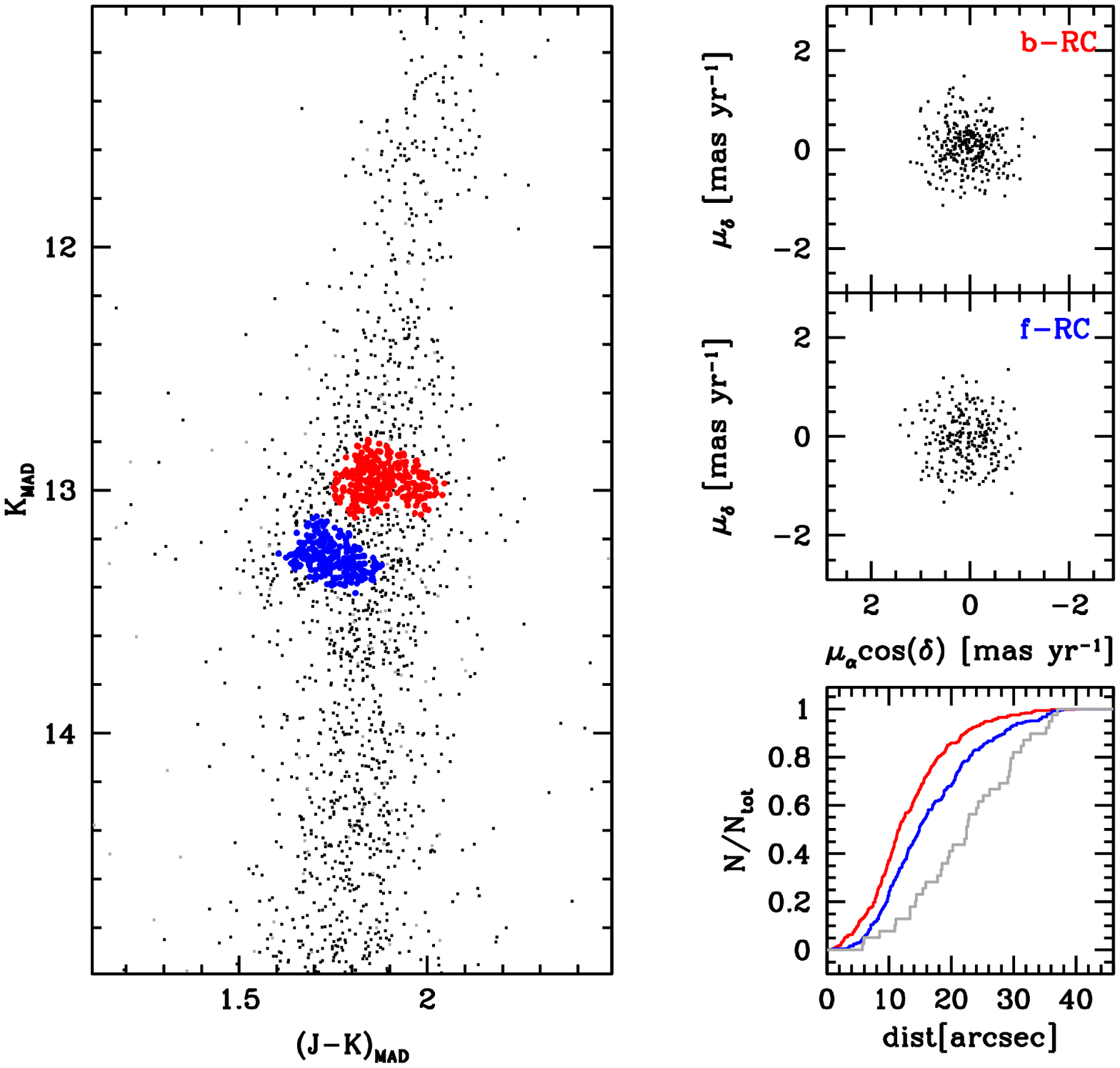}
 \caption{\small {\it Left panel:} IR MAD CMD of Terzan 5 zoomed in
   the RC region. Faint-RC stars are plotted in blue, while bright-RC
   stars are plotted in red. Stars rejected according to their PMs are
   shown in grey.  {\it Upper- and central-right panels:} VPDs of the
   bright-RC and faint-RC sub-samples, respectively: the stars of both
   populations lie within the 1.5 \masyr\ circle adopted as membership
   selection criterion. No clearcut difference between the two
   distributions is visible. {\it Lower-right panel}: cumulative
   radial distributions of bright-RC, faint-RC and likely non-members
   stars selected in the IR CMD (red, blue and grey lines,
   respectively). Clearly, non-member stars are less
   concentrated than cluster members, as is expected for
   field sources.}
\label{vpd2pops}
\end{figure}

Figure \ref{vpd2pops} highlights the CMD and VPDs for sub-samples of
stars properly selected in the two RCs. As can be seen, the two PM
distributions appear quite symmetric, both showing a small (0.5
\masyr) dispersion around the origin. While these PMs cannot be used
to reveal possible intrinsic kinematical differences between the two
populations, we can solidly conclude that these stars are all 
members of Terzan 5.

Overall the number of contaminants in the IR CMD (selected as stars
outside a distance of 1.5 \masyr\ from the center of the VPD) is much smaller
than what observed in the optical plane. This is because the MAD
photometry corresponds to a smaller FoV (with respect to the ACS one)
and to the very central region of the system, where the cluster
population is expected to dominate. It is also worth noticing that
both the RC sub-samples are much more centrally concentrated than the
likely Galactic field contaminants selected at the same magnitude
level (see the red, blue and grey lines, respectively, in
Fig.\ref{vpd2pops}). According to a Kolmogorov-Smirnov test performed
on the three samples, the probability of PM-rejected stars to have
been extracted from one of the two RC populations is always smaller
than 10$^{-4}$. This further confirms that stars rejected according to
our PM-membership selection are field objects, homogeneously
distributed in the sampled FoV.

In conclusion, the relative PMs measured in this work and applied to
the MAD CMD definitely demonstrate that {\it the two major
  sub-populations photometrically discovered in Terzan 5 are both
  members of the system}.

\subsection{Spectroscopic targets}
In \citet{massari14b} we presented the distribution of the iron
abundance for the largest sample of stars in Terzan 5. In that work,
target membership was inferred from radial velocities and the fraction
of contaminating stars was estimated statistically from the chemical
and kinematical characterization of the field surrounding the system
itself (\citealt{massari14a}). According to that work, in the
innermost 200\arcsec the expected fraction of contaminating field
stars is about 16\% (18 out of 114 stars).

\begin{figure}[!htb]
 \centering%
 \includegraphics[scale=0.5]{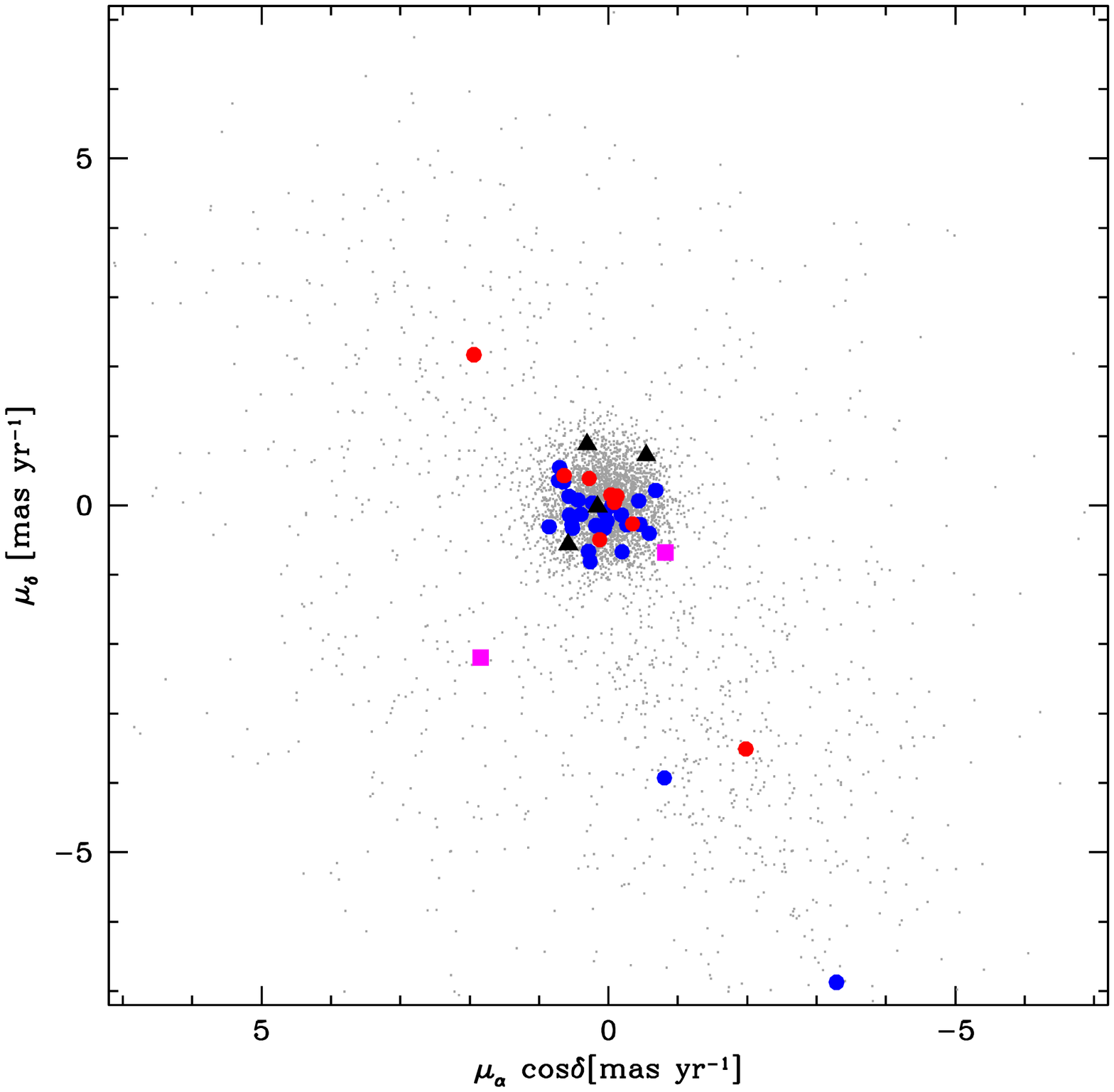}
 \caption{\small VPD for the spectroscopic targets discussed in
   \cite{massari14b} in common with our PM catalog. Blue circles are
   stars with $-0.6\le$[Fe/H]$<0$, while red circles have
   $0\le$[Fe/H]$<0.5$. These represent to the two main populations of
   Terzan 5. Black filled triangles correspond to stars belonging to
   the newly discovered third, metal-poor population. All the three
   components are clearly genuine members of the system, being located
   well within the PM distribution of cluster members. The two magenta
   filled squares represents the super metal-rich stars of the
   spectroscopic sample and, at odds with the other sources, they seem
   not to belong to Terzan 5.}
\label{vpd_targ}
\end{figure}

The measure of relative PMs gives the great opportunity to verify the
membership of each single star in the ACS FoV (instead of adopting a
statistical approach only) and to check whether the estimate of the
contribution from contaminating stars is reliable. To this aim,
we cross-correlated them with the iron abundance catalog.  We found
only 42 out of 135 targets in common, because the majority of the spectroscopic
targets are either bright RGB stars, which saturate even in the short
exposures, or stars located outside the ACS FoV. 
The location of these targets in the VPD is shown in Figure
\ref{vpd_targ}, where blue circles mark stars with
$-0.6\le$[Fe/H]$<0$, while red circle correspond to more metal rich
objects ($0\le$[Fe/H]$<0.5$). Clearly, with only a few exceptions (4
targets), all spectroscopic targets appear to be members of the
system, having PMs (and radial velocities) well within the
characteristic cluster distribution.  Also four stars belonging to the
newly discovered metal-poor component of Terzan 5 (with
[Fe/H]$\simeq-0.8$, see \citealt{o13, massari14b}) have measured PMs
and are shown in the figure (black triangles). As apparent, they are
also members of Terzan 5. Finally, the two magenta filled squares
correspond to the super metal-rich stars ([Fe/H]$>0.5$) found in
\cite{massari14b} and suggested to be non-members. Indeed, their PMs
are either outside the distribution of member stars, or just at its
edge.

Overall, 6 out of 42 spectroscopic targets appear to be field
stars. Such a fraction corresponds to the 14\% of the sample, in very
good agreement with the statistical estimate obtained from our
previous spectroscopic studies. Therefore, these results confirm that
the iron distribution of Terzan 5 shown in \cite{massari14b} is made
up of genuine cluster members, showing a huge internal iron spread of
more than 1 dex.

\subsection{Absolute proper motion}
One of the most intriguing application of PM measures is the
determination of the absolute motion of a stellar system. The best way
is to relate the mean motion of all the stars in the
system\footnote{Given the large number of measured PMS, this kind of
  analysis usually does not require the accuracy needed to study the
  internal kinematic of a system.} to the absolute reference frame
defined by very distant objects, such as background galaxies or
quasars, which appear as zero-motion sources (see, e.g.,
\citealp{dinescu99, mahmud, bellini10, lepine11, sohn12, massari13}).
Unfortunately, because of the high stellar density and the 
very large extinction in the direction
of Terzan 5, we have not been able to detect such objects in the ACS
FoV.

We then cross-correlated our PM catalog with many public catalogs of
stellar PMs, such as NOMAD (\citealt{zacharias}), the Guide Star
Catalog (GSC) version 2.3, UCAC4 (\cite{ucac4}) and the Yale/San Juan
Southern Proper Motion catalog (\citealt{platais},
\citealt{dinescu97}). However, since these catalogs have PMs measured
only for very bright sources, we found just fours stars in common with
our PM sample, all having PM uncertainties larger than 8 \masyr\ in
the public data sets. Hence, no meaningful result about the absolute
PM of Terzan 5 can be obtained from this kind of analysis.

An interesting alternative is described in (\citealt{ortolani11}, see also
\citealt{rossi15}), where the authors anchor the motion of the GC HP1 
to that of the underlying
bulge population. The bulge PM is the composition of its internal
kinematics and the reflex motion of the Local Standard of Rest
(LSR).  In the direction of Terzan 5, the radial velocity distribution
of bulge stars peaks at $v_{\rm rad}=21.0\pm4.6$ km s$^{-1}$
\citep{massari14a}. Assuming that bulge stars are at the same distance
of the Galactic center (here we adopt 8.4 kpc following
\citealt{ghez08, ortolani11}, but see also \citealt{marel12} for a
complete overview of the topic), the tangential component of the bulge
velocity is $v_{\rm tan}=v_{\rm rad}\sin(l)/\cos(l)=1.41\pm0.31$ km
s$^{-1}$, or in Galactic coordinates $\mu_{l}^{\rm bulge}=0.05\pm0.01$
\masyr. Since the bulge shows cylindrical rotation \cite{howard08,
  kunder12, ness13kin, zoccali14}, we can assume $\mu_{b}^{\rm
  bulge}=0$ \masyr. By summing these values to the motion of the LSR
($\mu_{l}\cos(b)=6.10\pm0.25$ \masyr, $\mu_{b}=0$ \masyr,
corresponding to $v_{\rm LSR}=243\pm10$ km s$^{-1}$; see
\citealp{ortolani11}), we obtain the total motion of the bulge at
Terzan 5 coordinates: ($\mu_{l}\cos(b)$, $\mu_{b})_{\rm
  bulge}=(6.15\pm0.25$, 0) \masyr.

As a second step for determining the PM of Terzan 5 relative to that
of the bulge, we excluded foreground disk contaminants belonging to
the blue plume in the CMD by selectining a sub-sample of stars with
({\it m}$_{{\rm F606W}}$-{\it m}$_{{\rm F814W}})>3.9$ and {\it
  m}$_{{\rm F606W}}<24.5$, which do not saturate even in the long
exposures. We also applied a selection in terms of PM errors, by
excluding stars with uncertainties larger than 0.3 \masyr\ in the two
PM components. We then defined as Terzan 5 members all the stars
located within 2\masyr\ from the origin of the VPD.

\begin{figure}[!htb]
 \centering%
 \includegraphics[scale=0.5]{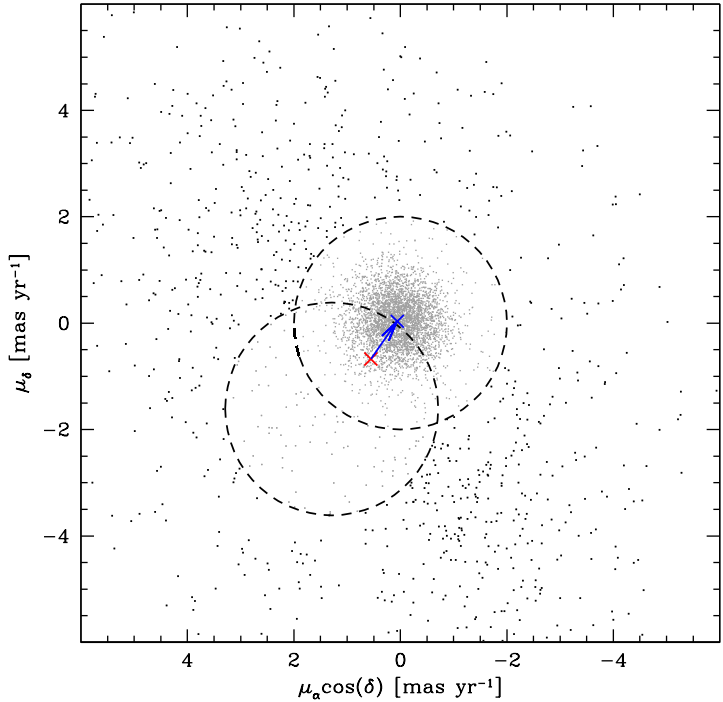}
 \caption{\small VPD of the sample of stars used to estimate the
   absolute PM of Terzan 5.  The black points correspond to the bulge
   stars surviving the last step of the selection procedure and
   effectively used as reference; the gray dots mark the Terzan 5
   member stars.  The mean motion of the bulge sample is marked with a
   red cross. The PM of Terzan 5 relative to the mean bulge motion is
   marked with a blue arrow.}
\label{absol}
\end{figure}

To select the sample of bulge stars we followed an iterative procedure
similar to that described by \cite{jay10} for the determination of the
center of $\omega$ Centauri.  After the exclusion of all Terzan 5
members, we computed the first guess mean motion of bulge stars. We
then excluded all objects contained within a circle of 2\masyr\ radius
positioned at the same distance as the ``Terzan 5 circle'' from the
first guess mean, on the symmetrically opposite side in the VPD. A new
value of the mean motion has been evaluated from this sample, and the
procedure has been repeated until convergence was reached.  Once the
Terzan 5 and the bulge samples are defined, we determined their
weighted mean motion with a 3-$\sigma$ clipping procedure that, after
the last step, included 3853 sources for the Terzan 5 sample and 797
for that of the bulge (black dots in Figure \ref{absol}). We computed
the uncertainty in the weighted mean motions as the standard
error-in-the-mean, i.e., the dispersion of the surviving stars around
the mean PM, divided by the square root of their number.

The resulting motion of Terzan 5 relative to the bulge (see the blue arrow
in Figure \ref{absol}) expressed in Galactic
coordinates (see \citealt{ortolani11} for the details of the
transformations) turned out to be $(\mu_l\cos(b)$, $\mu_b)_{\rm
  T5-bulge}=(0.26\pm0.10$, $0.83\pm0.12)$ \masyr\ , where the final 
uncertainties are the sum in quadrature of those computed for the two separated 
samples.
By subtracting the motion of the bulge quoted above, we finally
obtained the absolute PM of Terzan 5: ($\mu_l\cos(b)$, $\mu_b)_{\rm
  T5}=(-5.89\pm0.27$, $0.83\pm0.12)$ \masyr, where the total
uncertainties are the sum in quadrature between that coming from the
motion of the bulge and that associated to the motion of the cluster
relative to it.  This is the first absolute PM estimate ever obtained
for Terzan 5.

\subsection{The cluster orbit}

\begin{figure}
\centering
\includegraphics[width= 8.9 truecm]{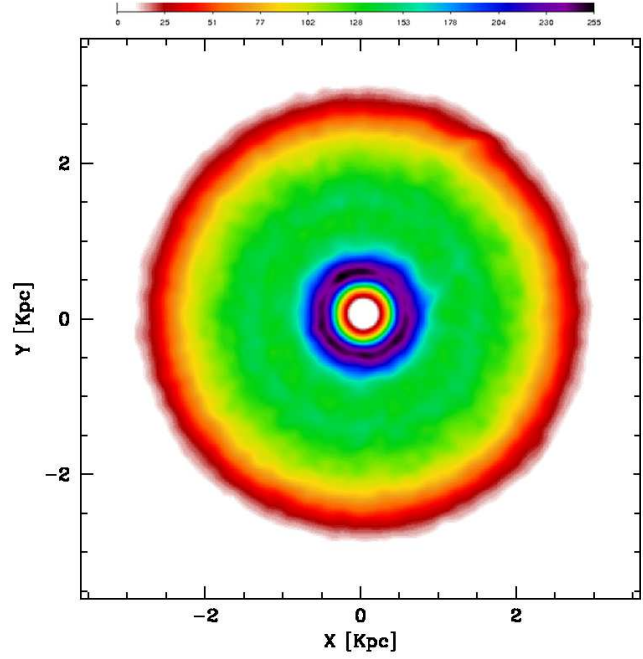}
\caption{Probability density in the equatorial Galactic plane of 1000
  simulated orbits of Terzan 5 time-integrated backwards for 12 Gyr.
  Darker colors correspond to larger probabilities.
\label{planexy}}
\end{figure}

\begin{figure}
\centering
\includegraphics[width= 8.4 truecm]{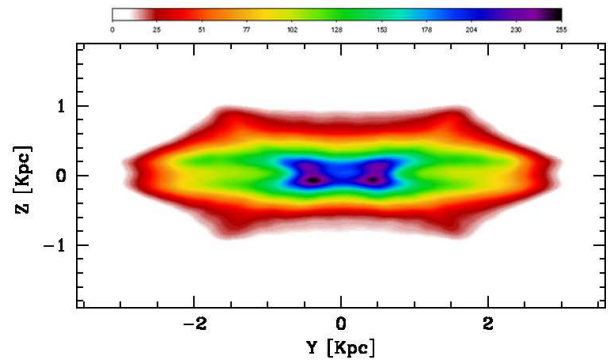}
\caption{As in Fig. \ref{planexy}, but for the orbit projection in
  the meridional Galactic plane.
\label{planerz}}
\end{figure}

On the basis of the absolute PM estimate derived in the previous
section, we performed a numerical integration of the orbit of Terzan 5
in the Galactic potential. We used the 3-component (bulge, disk and
halo) axisymmetric model from \cite{allen91}, which has been
extensively used and discussed in the literature to study orbits and
dynamical environmental effects on Galactic stellar systems
\citep[e.g.][]{allen06,montuori07,ortolani11,moreno14,zonoozi14},
thanks to its relative simplicity and fully analytic nature. We
adjusted the various model parameters to make the rotation velocity
curve match the value of 243 km s$^{-1}$ measured at the Solar
Galactocentric distance of 8.4 kpc (see above). The cluster PM were
reported in the cartesian Galactocentric reference frame, resulting in
a velocity vector $(v_x,v_y,v_z)=(-60.4\pm 1.3$, $85.7\pm 11.1$,
$35.0\pm 6.0)$ km s$^{-1}$ at the position $(x,y,z)=(-2.51\pm 0.30$,
$0.39\pm 0.02$, $0.17\pm 0.01)$ kpc. We adopted the convention in
which the $X$ axis points opposite to the Sun (i.e., the Sun position
is $(-8.4,0,0)$). The orbit was then time-integrated backwards for 12
Gyr, starting from the given initial (current) conditions and using a
2nd order leap-frog integrator \citep[e.g.][]{hockney} with a rather
small and constant timestep ($\sim 100$ kyr, corresponding to $\sim
1/300$ of the dynamical time at the Sun distance, computed as
$(R_0^3/GM_g)^{1/2}$, where $M_g\sim 1.7\times 10^{11} M_\odot$ is a
characteristic Galactic mass parameter). During the $>10^5$ timesteps
used to describe the entire orbit evolution, the errors on the
conservation of both the energy and the $Z$ component of angular
momentum were kept under control and never exceeded 1 part over $10^5$
and $10^{13}$, respectively.  To take into account the uncertainties
on the kinematic data, we generated a set of 1000 orbits starting from
phase-space initial conditions normally distributed within a $3\sigma$
range around the cluster velocity vector components and current
position, with $\sigma$ being equal to the quoted uncertainties on
these parameters. For all these orbits we repeated the backward
time-integration.  The probability densities of the resulting orbits
projected on the equatorial and meridional Galactic planes are shown
in Figures \ref{planexy} and \ref{planerz}, respectively. Darker
colors correspond to more probable regions of the space, i.e., to
Galactic coordinates crossed more frequently by the simulated
orbits. As apparent, the larger distances reached by the system during
its evolution are $R=3.5$ kpc and $|Z|=1.6$ kpc at a $3\sigma$ level
of significance, which roughly correspond to a region having the 
current size of the bulge.

In a forthcoming study, we plan to quantify the possible influence of
a central non axisymmetric component (the bar and/or the spiral
arms). Indeed, although triaxial potentials are expected to facilitate
the onset of very radial orbits (because of the lack of conservation
of any angular momentum component) with a possible prolongation of the
system orbital motion in much outer regions, a recent study by
\citet{moreno14} demonstrated (using a sophisticated and realistic
model) how this intuitive picture could be wrong, due to the complex
interplay between the gravitational potentials generated by the
various components. The relatively small perigalactic distance ($<500$
pc) that Terzan 5 could have achieved during its orbit could imply a
significant mass loss due to tidal erosion and thus represents a
further strong motivation for studying in more detail the dynamical
history of this system.
 
Based on the current results, we can conclude that the integration of
Terzan 5 orbit performed by means of this simple axisymmetric Galactic static
model suggests an in-situ formation for the cluster within the Galaxy, rather than
an external accreted origin. This supports the interpretation \citep{f09} of this system
as the remnant of one of the massive stellar clumps that may have
contributed to form the Galactic bulge (\citealt{noguchi99,
  immeli04}).

\section{CONCLUSIONS}
\label{conclusions}
By using high-resolution ACS/WFC images, we have measured relative PMs
in the direction of the stellar system Terzan 5.  We have been able to
separate cluster members from foreground disk stars and background
bulge contaminants. The PM-selected CMD clearly confirmed that the two
main populations with distinct iron content discovered in Terzan 5 are
genuine members of the system, thus putting to rest this issue which
had remained a matter of debate. Moreover, the measured PMs demonstrate that also the
third, metal-poorer and $\alpha$-enhanced population clearly belongs
to the system. Our findings therefore confirms that Terzan 5 is,
together with $\omega$ Centauri (\citealt{omega, pancino2000,
  origlia03, jp10, villanova14}), the stellar system with the largest
internal iron spread in the Galaxy.

Finally, by comparing the motion of Terzan 5 members to that of bulge
stars, we estimated for the first time the absolute PM of this system,
finding ($\mu_{l}\cos(b)$, $\mu_{b})_{\rm T5}= (-5.89\pm0.10$,
$0.83\pm0.12)$ \masyr. 
The backward integration of its orbit, computed under the
assumption of an axisymmetric Galactic model, provides
another indication that supports the scenario according to which Terzan 5 is the
remnant of one of the primordial stellar clumps that may have
contributed to form the galactic Bulge, and not an object accreted
from outside the Milky Way.

\acknowledgments We thank the anonymous Referee for his/her useful 
comments which helped us to improve our paper. 
The authors are also grateful to Dr.  Lucie J\'ilkov\'a for her valuable
suggestions on the cluster orbit integration.
This research is part of the project {\it Cosmic-Lab}
(web site: http://www.cosmic-lab.eu) funded by the European Research
Council (under contract ERC-2010-AdG-267675). 
D.M. thanks the Kapteyn Astronomical Institute for hospitality.
R.M.R. acknowledges support from AST-1413755 from the National Science
Foundation. The techniques applied in the present work have been developed in the
context of the HSTPROMO collaboration\footnote{For details see
  HSTPROMO home page at http://www.stsci.edu/~marel/hstpromo.html}
(\citealt{hstpromo, bellini14}), which aims at improving our
understanding of the dynamical evolution of stars, stellar clusters
and galaxies in the nearby Universe through the measurement and
interpretation of PMs.

\end{document}